\documentclass[a4paper,11pt]{article}
\usepackage[utf8x]{inputenc}

\usepackage{amsmath}
\usepackage{amsfonts}
\usepackage{amssymb}

\usepackage{todonotes}

\usepackage[comma]{natbib}

\usepackage[ruled]{algorithm2e}

\usepackage{geometry}

\usepackage{microtype}
\usepackage{subfig}

\usepackage{xcolor}

\newcommand{\Cor}{\operatorname{Cor}}

\title{Generalized Gaussian Random Fields\\ using hidden selections}
\author{
       Kjartan Rimstad \& Henning Omre \vspace{0.15cm} \\
\small Department of Mathematical Sciences\\
\small Norwegian University of Science and Technology\\
\small Trondheim, Norway \vspace{0.35cm} \\
Running head: Selection Gaussian Random Fields 
}

\begin{document}

\maketitle

\begin{abstract}
We study non-Gaussian random fields constructed by the selection normal distribution, and we term them selection Gaussian random fields. The selection Gaussian random field can capture skewness, multi-modality, and to some extend heavy tails in the marginal distribution.  We present a Metropolis-Hastings algorithm for efficient simulation of realizations from the random field, and a numerical algorithm for estimating model parameters by maximum likelihood. The algorithms are demonstrated and evaluated on synthetic cases and on a real seismic data set from the North Sea. In the North Sea data set we are able to reduce the mean square prediction error by $20$-$40$\% compared to a Gaussian model, and we obtain more reliable prediction intervals.
\end{abstract}

\vspace{0.5cm} 


\clearpage

\section{Introduction}

Statistical spatial prediction is an important problem in many earth science and engineering applications, such as petroleum exploration, mining, hydrology, and meteorology. The variables of interest are often considered to be a realization from a random field, and we want to predict the variable in unobserved parts of the random field by exploiting the dependence structure of the random field. Most prediction methods assume, explicitly or implicitly, that the observations come from one realization of a Gaussian random field, and use of optimal linear predictors lead to the commonly used kriging technique \citep{cressie93}. However, many data sets from natural sciences have non-Gaussian characteristics, such as skewness, multi-modality, and/or heavy tails. 

The non-Gaussian effects are often reduced by transforming the field by a non-linear transformation into an approximately Gaussian random field, for example by using the Box-Cox family of power transformation \citep{Box1964,Diggle2007}. The transformation parameters are usually unknown and have to be estimated, which may be problematic. One example of this approach is presented in \citet{DeOliveira1997} where a transformed Gaussian random field is considered in a Bayesian setting. An alternative strategy is to assume that the random field is a non-Gaussian random field that captures skewness, multi-modality, and/or heavy tails. The latter approach is chosen in this study.

We consider multivariate probability distributions which are constructed by modifying a symmetric probability density function (pdf). The idea of modifying symmetric probability densities of a random variable is made popular by \citet{Azzalini1985}, who introduced the skew-normal distribution. Later, the family of skew-normal distributions is extended to the multivariate skew-normal distribution in \citet{Azzalini1996}. Several authors have generalized these distributions, and a summary is presented in \citet{Arellano-Valle2006a}. The book edited by \citet{Genton2004} provides a detailed overview of these distributions.

In this study we consider a family of distributions arising from applying various forms of selection mechanisms on symmetric distributions as discussed in \citet{Arellano-Valle2004} and \citet{Arellano-Valle2006a}. We work in a spatial setting and define random fields constructed by using this family of distributions, which we term selection Gaussian random fields. In \citet{Kim2004}, \citet{Allard2007}, and \citet{Rimstad2012} simple selection distributions are used to define skew-Gaussian random fields, but it appears as difficult to model high degree of skewness with these random field due to correlation effects, see \citet{Rimstad2012}. In the current study we generalize these random fields by using more general selection mechanisms and use these distributions to define random fields. We are then able to model skewness, multi-modality, and to some extent heavier tails. The selection Gaussian random field may also be seen as an alternative to spatial mixture models for modeling multi-modality in 
random field, for example by using latent discrete Markov random field models \citep[see e.g.][]{besag1974,Kaiser2002}.

Selection mechanisms can be applied to any distribution but we only consider the multivariate selection normal distribution because this model inherits important properties from the multivariate normal distribution, such as being closed under marginalization, conditioning, and linear transformation. The closure properties and the relation to the multivariate normal distribution are important because they simplify sampling and inference algorithms, and spatial prediction.

In this study we define selection Gaussian random fields, and we generalize the Metropolis-Hasting algorithm for sampling and the Monte Carlo maximum likelihood parameter estimation algorithm in \citet{Rimstad2012} to be applicable for selection Gaussian random fields. We demonstrate sampling, inference, and prediction by synthetic examples. Lastly we use a multivariate selection Gaussian random field in a predictive setting on a real seismic data set from the North Sea.

\section{Model}


The multivariate selection normal distribution, defined in \citet{Arellano-Valle2004} and \citet{Arellano-Valle2006a}, extends the multivariate normal distribution to allow modeling skewness, multi-modality, and to some extent heavy tails, while retaining many important properties of the normal distribution. Let the random vector $\mathbf U$ be multivariate normal distribution by using the notation:
\begin{align}
\mathbf U 
=
\left(
\begin{array}{c}
\mathbf U_1 \\
\mathbf U_2
\end{array}
\right)
\sim
N_{p+q} 
\left[
\boldsymbol \mu = 
\left(
\begin{array}{c}
 \boldsymbol \mu_1 \\
 \boldsymbol \mu_2
\end{array}
\right)
,
\boldsymbol \Sigma =
\left(
\begin{array}{cc}
 \boldsymbol \Sigma_1 & \boldsymbol \Sigma_{12} \\
 \boldsymbol \Sigma_{21} & \boldsymbol \Sigma_2 \\
\end{array}
\right)
\right],
\end{align}
where $\mathbf U \in \mathbb{R}^{p+q}$, $\mathbf U_1, \boldsymbol \mu_1 \in \mathbb{R}^p$, $\mathbf U_2, \boldsymbol \mu_2 \in \mathbb{R}^q$, $\boldsymbol \Sigma_1 \in \mathbb{R}^{p\times p}$, $\boldsymbol \Sigma_2 \in \mathbb{R}^{q\times q}$, $\boldsymbol \Sigma_{12} = \boldsymbol \Sigma_{21}^T \in \mathbb{R}^{p\times q}$, $T$ denotes matrix transpose, and $N_{n}(\boldsymbol \mu, \boldsymbol \Sigma)$ denotes $n$-dimensional multivariate normal distribution with mean vector $\boldsymbol \mu$ and covariance matrix $\boldsymbol \Sigma$. Then  $\mathbf X = [\mathbf U_1 \mid \mathbf U_2 \in A]$ is multivariate selection normal  distributed, with respect to an arbitrary set $A \subseteq R^q$, denoted $SLCT\mathrm{-N}_{p,q}(\boldsymbol \mu, \boldsymbol \Sigma , A)$. The corresponding pdf is
\begin{align}
&f_{p,q} (\mathbf x; \boldsymbol \mu, \boldsymbol \Sigma, \mathbf A) 
=
\phi_p(\mathbf x; \boldsymbol \mu_1, \boldsymbol \Sigma_1)  \;
\frac{
\Phi_q(A; \boldsymbol \mu_2 + \boldsymbol  \Sigma_{21} \boldsymbol  \Sigma_1^{-1} ( \mathbf x - \boldsymbol \mu_1), \boldsymbol \Sigma_2 - \boldsymbol \Sigma_{21} \boldsymbol  \Sigma_1^{-1} \boldsymbol \Sigma_{12} )
}{
\Phi_q(A; \boldsymbol \mu_2, \boldsymbol \Sigma_2)
}, \label{eqn:density}
\end{align}
where $\phi_n(\mathbf x;\boldsymbol \mu,\boldsymbol \Sigma)$ is the $n$-dimensional multivariate normal pdf with mean vector $\boldsymbol \mu$ and covariance matrix $\boldsymbol \Sigma$, and $\Phi_n (A; \boldsymbol \mu, \boldsymbol \Sigma) = p(\mathbf Y \in A)$ for $\mathbf Y \sim N_n(\boldsymbol \mu, \boldsymbol \Sigma) $. The latter corresponds to the probability for a normally distributed variable $\mathbf Y$ with expectation $\boldsymbol \mu$ and variance $\boldsymbol \Sigma$ to be in the set $A$. 

The properties of the multivariate selection normal distribution are presented in \citet{Arellano-Valle2006a}, and it is shown that the multivariate selection normal distribution inherits important properties from the multivariate normal distribution, such as being closed under marginalization, conditioning, and linear transformation.

In the current study we consider the multivariate selection normal distribution in a spatial setting; thus we define a spatial random field based on the multivariate selection normal distribution. Let $\left\lbrace Z(\mathbf s): \mathbf s \in \mathcal D \subseteq \mathbb{R}^d \right\rbrace$ be a random field of real-valued variables, where $\mathcal D$ is a spatial set of dimension $d$ and $\mathbf s \in \mathbb R^d$ is a generic location in $\mathcal D$. Then the random field $\left\lbrace Z(\mathbf s): \mathbf s \in \mathcal D \right\rbrace$ is a Gaussian random field if for all configurations of points $\mathbf s_1,\ldots, \mathbf s_n$ and all $n>0$ the pdf of $\mathbf Z = [Z(\mathbf s_1), \ldots, Z(\mathbf s_n)]^T$ is multivariate normal.

The selection Gaussian random field is defined by considering the bivariate Gaussian random field 
\begin{align}
\left\lbrace 
\mathbf U(\mathbf s)
=
\left(
\begin{array}{c}
U_1(\mathbf s) \\
U_2(\mathbf s)
\end{array}
\right): \ \ 
\mathbf s \in \mathcal D
\right\rbrace.
\label{eqn:bivariate}
\end{align}
The fixed configuration $\mathbf s'_1, \ldots,\mathbf s'_q$, with fixed finite $q$, defines $\mathbf U_2 = \left[ U_2(\mathbf s'_1), \ldots, U_2(\mathbf s'_q) \right]$. For a specified set $A \subseteq \mathbb R^q$ we define $\left\lbrace X(\mathbf s) =  \left[ U_1(\mathbf s) \mid \mathbf U_2 \in A \right] : \mathbf s \in \mathcal D \right\rbrace$, which is a selection Gaussian random field if for all configurations of points $\mathbf s_1,\ldots, \mathbf s_p$ and all $p>0$ the pdf of $\mathbf X = [X(\mathbf s_1), \ldots, X(\mathbf s_p)]^T$ is multivariate selection normal distributed. Or equivalent, if the Gaussian random field $U_1(\mathbf s)$ and $\mathbf U_2$ are jointly Gaussian, then $\left\lbrace X(\mathbf s) =  \left[ U_1(\mathbf s) \mid \mathbf U_2 \in A \right] : \mathbf s \in \mathcal D \right\rbrace$ is a selection Gaussian random field.

A special case occurs if  $U_1(\mathbf s)$ and $U_2(\mathbf s)$ in Expression \ref{eqn:bivariate} are independent, then $X(\mathbf s)$ is a Gaussian random field. Moreover, if $|\Cor(U_1(\mathbf s),U_2(\mathbf s))| = 1$ for all $\mathbf s \in \mathcal D$, then $X(\mathbf s)$ is a truncated Gaussian random field. When $U(\mathbf s)$ is a stationary Gaussian random field and the discretization $\mathbf s'_1, \ldots,\mathbf s'_q$ is a regular grid over $\mathcal D$, then the marginal pdfs of $X(\mathbf s)$ is stationary in the discretization locations $\mathbf s'_1, \ldots,\mathbf s'_q$ when border effects caused by finite $\mathcal D$ are ignored. 

In the current study we consider the family of distributions where the set $A \subseteq \mathbb R^q$ is on the form $\left\lbrace \mathbf y \in A \right\rbrace  =  \left\lbrace y_i \in A_i, \; i = 1,\ldots,q  \right\rbrace $, with $A_i = \cup_{j=1}^{m} [a_{ij},b_{ij}] $. Hence $A_i \subseteq \mathbb R^1$ may consist of several line segments of $\mathbb R^1$. A special case occurs if the selection sets are $A_i = (-\infty, 0], \ i = 1,\ldots q$, then the random field $X(\mathbf s)$ is a closed skew normal (CSN) random field as defined in \citet{Allard2007}.

We consider selection Gaussian random fields on a regular grid with stationary parameters, with pdf given in Expression \ref{eqn:density}. A simple model with few parameters is used such that we are able to make parameter inference from one realization of the random field. The model should however be sufficiently flexible to exhibit non-Gaussian properties. We use $q = p$, $\boldsymbol \mu_1 = \mu \mathbf 1$, and we let $\boldsymbol \mu_2 = \mathbf 0$, where $\mathbf 0 \in \mathbb R^p$ and $\mathbf 1 \in \mathbb R^p$ are vectors of zeros and ones, respectively. Note that the design of the set $A \subseteq \mathbb R^q$ also is considered to be a model parameter in the random field; hence we need to have $\boldsymbol \mu_2 = \mathbf 0$ to make the model identifiable. The covariance structure is defined to be on a form similar to the one in \citet{Rimstad2012}:
\begin{align}
\boldsymbol \Sigma 
&=
\left[
\begin{array}{cc}
\sigma^2 \mathbf C 		& \gamma \sigma \mathbf C \\
\gamma \sigma \mathbf C	& (1-\gamma^2) \mathbf I_p + \gamma^2  \mathbf C
\end{array}
\right], \label{eqn:covariance}
\end{align}
where $\sigma^2$ is a scale parameter, $|\gamma| \leq 1$ is a coupling parameter between the observed and the truncated random fields, $\mathbf I_p$ is a $p$-dimensional identity matrix, and $\mathbf C$ is a correlation matrix with an exponential correlation function $\rho(\mathbf x', \mathbf x'') = \mathrm{exp} \{ - \left| x'_1- x''_1 \right|^2/d_h^2 - \left| x'_2 - x''_2 \right|^2/d_v^2 \} $ where $d_h$ and $d_v$ are horizontal and vertical range parameter, respectively. Expression \ref{eqn:density} then becomes
\begin{align}
f_{p,q} (\mathbf x; \mu, A, \sigma^2, \gamma, d_h, d_v) 
&=
\phi_p(\mathbf x; \mu \mathbf 1, \sigma^2 \mathbf C)  \;
\frac{
\Phi_q( A;\frac{\gamma}{\sigma} (\mathbf x - \mu \mathbf 1), (1-\gamma^2) \mathbf I_p)
}{
\Phi_q( A;\mathbf 0, (1-\gamma^2) \mathbf I_p + \gamma^2 \mathbf C)
} \notag \\
&=
\phi_p(\mathbf x; \mu \mathbf 1, \sigma^2 \mathbf C)  \;
\frac{
\prod_{i=1}^q \Phi_1(A_i; \frac{\gamma}{\sigma} ( x_i - \mu ), 1-\gamma^2 )
}{
\Phi_q( A; \mathbf 0, (1-\gamma^2) \mathbf I_p + \gamma^2 \mathbf C)
}
. \label{eqn:param_density}
\end{align}
The parameters in the model are $\mu$, $\sigma^2$, $\gamma$, $d_h$, $d_v$, and the design of the set $A $. The constraints $\sigma^2,d_h,d_v > 0$ and $|\gamma| \leq 1$ ensure that $\boldsymbol \Sigma$ is positive semidefinite and hence a valid covariance matrix. 

The stochastic expression of the discretized selection Gaussian random field $\mathbf X$ is
\begin{align}
\mathbf  X
&= [\mathbf U_1 \mid \mathbf  U_2 \in A] \notag  \\
&=
\mu \mathbf 1 + \gamma \sigma \mathbf C ((1-\gamma^2) \; \mathbf I_p + \gamma^2 \mathbf  C)^{-1} [\mathbf U_2|\mathbf U_2 \in A] +  \mathbf V,
\end{align}
with $\mathbf V \sim N(\mathbf 0, \sigma^2 \mathbf  C  - \gamma^2  \sigma^2 \mathbf C  ((1-\gamma^2) \; \mathbf I_p + \gamma^2 \mathbf C)^{-1} \mathbf C )$. Expressions for the mean and covariance matrix of $\mathbf X$ can be found in \citet{Arellano-Valle2006a}, but they are in general not easy to evaluate without simulation due to the $[\mathbf U_2|\mathbf U_2 \in A]$ term.

\vspace{0.5cm}

\begin{table}
\begin{center}
\begin{small}
\begin{tabular}{c|ccccl }
Case & $\gamma$ & $d_h$ & $d_v$ & $A_i$ & description \\ \hline
1    & 0.8000   & 2.0   & 2.0  & $(\infty, -0.3] \cup [0.3, \infty)$ & sym. bimodal iso. \\ 
2    & 0.6500   & 6.0   & 0.85  & $(\infty, -0.3] \cup [0.3, \infty)$ & asym. bimodal aniso. \\
3    & 0.9250   & 2.0   & 0.60  & $(\infty, -0.85] \cup [0.8, \infty)$& sym. bimodal aniso. \\
4    & 0.9995   & 3.0   & 3.0  & $[-0.45, -0.2] \cup [-0.1, 0.1] \cup [0.2, 0.45]$ & sym. trimodal iso.\\
5    & 0.7000   & 2.0   & 2.0  & $(\infty, -0.7] \cup [-0.1, 2.5]$    & asym. unimodal iso. \\
6    & 0.7000   & 2.0   & 2.0  & $(\infty, -1.75] \cup [-0.5, 0.5] \cup [1.75, \infty)$ & sym. heavy tailed iso. \\
\end{tabular}
\end{small}
\end{center}
\caption{Model parameters for six cases, with $\mu = 0 $ and  $\sigma^2=1$ for all cases.}
\label{tbl:synthetic_param}
\end{table}

There are many possible parameterizations of the set $A$, and we explore six designs, all of them stationary models where $A_i$ is identical for all $i$. We wish to reproduce multi-modality, skewness, and to some extent heavy tails in the marginal distribution of the random field. The parameters for the various cases are summarized in Table \ref{tbl:synthetic_param}.

In order to simulate realizations from the selection Gaussian random field we generalize the Metropolis Hastings (MH) algorithm presented in \citet{Rimstad2012} by allowing more general selection sets $A$. The algorithm is summarized in Appendix \ref{sec:app_sample}. The algorithm is a block proposal MH-algorithm, and we normally use block sizes about $100$ which in our examples give an acceptance rate of about $0.25$. We sample $(64 \times 64)$ grid random fields, and the computer demand for generation of one realization is a couple of minutes on a regular laptop computer. The burn-in and mixing appear as satisfactory and are not displayed.

\begin{figure}
 \centering
 \includegraphics[height=0.925\textheight]{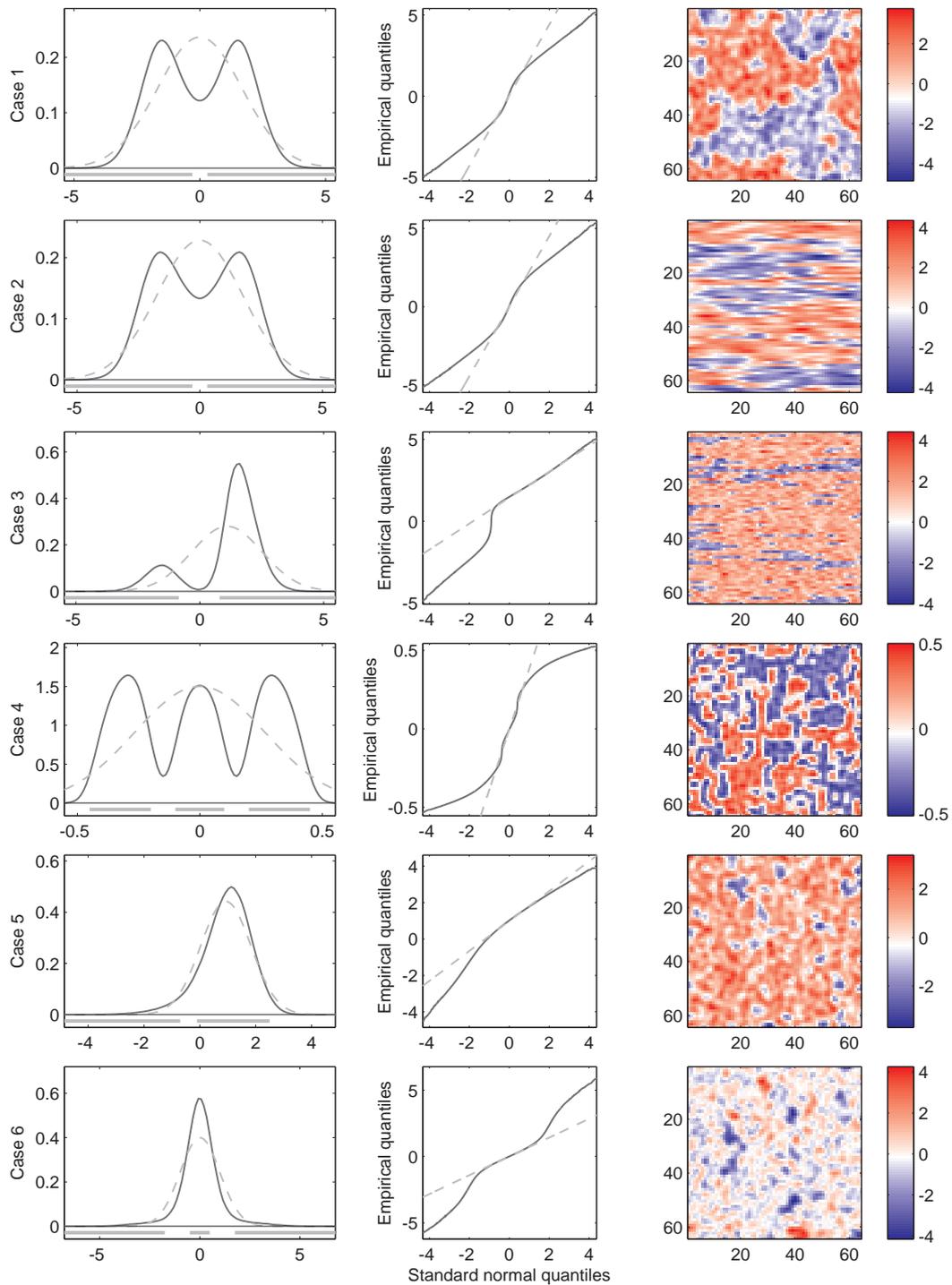}
 \caption{First column: marginal distribution of selection Gaussian random field is solid black, standard normal distribution is dashed gray, and feasible sets on latent random field on axis in solid gray. Second column: quantile-quantile plot of marginal selection Gaussian random field versus theoretical quantiles from the normal distribution. Third column: realization from selection Gaussian random field.}
 \label{fig:synthetic_marginals}
\end{figure}

Figure \ref{fig:synthetic_marginals} displays the results from the six cases. The first column displays the marginal distribution in the $(32, 32)$ location of the $(64 \times 64)$ grid random field compared with a univariate normal distribution with the same moments as the marginal selection normal distribution. The feasible set $A$ of the latent truncated random field $\mathbf U_2$ is illustrated with a thick gray line at the bottom of the display. The feasible set of the latent truncated random field $\mathbf U_2$ is comparable to the marginal of the selection Gaussian random field $\mathbf X$ because we have $\sigma^2=1$, otherwise we would have to correct for scaling. The second column displays quantile-quantile plots of the marginal distributions displayed in the first column versus quantiles from the standard normal distribution. The last column displays realizations from selection Gaussian random fields.

The first row in Figure \ref{fig:synthetic_marginals}, case 1, displays a symmetric bimodal spatially isotropic random field. The feasible region for the latent random field is absolute values greater than $0.3$. The marginal distribution is symmetric and bimodal, and the quantile-quantile plot shows clear deviations from the normal distribution. In the realization of the random field the two modes are visible as two separated levels with sharp transitions between them.

Case 2 is displayed in the second row in Figure \ref{fig:synthetic_marginals} and this random field is also symmetric and bimodal, but is a spatially anisotropic random field. The  feasible region for the latent random field is absolute values greater than $0.3$, as in case 1. In this case the horizontal spatial correlation is increased and the vertical one decreased, while the coupling parameter is reduced. The resulting random field is clearly layered, with marginal distribution very similar to case 1.

The third row in Figure \ref{fig:synthetic_marginals}, case 3, displays an asymmetric bimodal spatially anisotropic random field. In this case the truncation is asymmetric and further in the tails than in previous cases. The occurrence of two clearly separated modes are possible due to low spatial correlation, because this low correlation allows larger jumps. The asymmetric truncation cause the mode to the left to be smaller than the mode to the right. Even though the correlation in the field is low the random field has clear spatial anisotropic structure and two well separated modes. 

Case 4 is displayed in the forth row in Figure \ref{fig:synthetic_marginals} and is a symmetric trimodal spatial isotropic random field. In this case the truncated random field has three feasible symmetric intervals, which provides trimodal symmetric marginal distribution. Note also that this case has three closed feasible intervals with finite endpoints, compared to the previous cases with two feasible intervals with one infinite endpoint each; thus the tails are lighter for this random field. The three modes are well separated, and clearly visible in the realizations. The spatial transitions in the realization between values of the two outer modes seem to always pass through the middle mode, though.

Bi- and multi-modal models provide alternatives to spatial mixture models with for example a hidden discrete Markov random field model \citep[see e.g.][]{besag1974,Kaiser2002}. The current model has the advantage that it is easier to construct efficient simulation algorithm for it than for a discrete Markov random field model due to the current model's relationship to the normal distribution. Note also that the truncation region $[-0.3,  0.3]$ in case 2 and 3 is small, but the effects on the marginal distribution is substantial. This is caused by the spatial correlation effect that decreases the probability of the latent variables to be close to the truncation region. This effect is discussed in more detail for the CSN random field in \citet{Rimstad2012}. 

The fifth row in Figure \ref{fig:synthetic_marginals}, case 5, displays a skewed random field. The CSN random field considered in \citet{Allard2007} and \citet{Rimstad2012} only truncate one side of the hidden random field and the model formulation put constraints on the degree of skewness. In this case we introduce one additional truncation interval, which allows a more flexible skewness structure in the random field, as illustrated in Figure \ref{fig:synthetic_marginals}. The skewness is evident in the marginal distribution, in the quantile-quantile plot, and in the realization of the selection Gaussian random field.

The last row in Figure \ref{fig:synthetic_marginals}, case 6, displays a symmetric heavy tailed random field. We use symmetric truncation, and the idea is to force higher probability density around the mean and in the tails, which is visible in the marginal distribution in Figure \ref{fig:synthetic_marginals}. Note that the extreme tails still decays exponentially, see Expression \ref{eqn:param_density}, and the quantile-quantile plot, but the more likely visible effects of the heavy tails are apparent. The closest univariate Student-$t$ distribution is one with about $2$ degrees of freedom, if we ignore the extreme tails. We could alternatively have substituted the multivariate normal distribution with a multivariate t-distribution in the construction of the selection Gaussian random field \citep[see e.g.][]{Arellano-Valle2006a}, which can be done with only a small computational cost \citep{Genz2009}. By using the multivariate t-distribution we would get heavier tails in the marginal distribution, but each 
realization of the random field would look identical to the selection Gaussian random field up to a scaling factor; thus the parameters of the model would not be identifiable. The multivariate t-distribution also lacks some of the closure properties the multivariate normal distribution \citep[see e.g.][]{Roislien2006}. 

Case 1 to 6 illustrate some of the characteristics the selection Gaussian random field is able to model. We are able to generate random fields with multi-modality in the marginal distribution, symmetric and asymmetric marginal distributions, and light and to some extent heavy tails.

\section{Parameter estimation}

We follow \citet{Rimstad2012} and use a maximum likelihood approach to estimate the parameters with a Monte-Carlo approximated likelihood algorithm \citep{Geyer1992}. The same parameterization as in the previous section is used and we estimate parameters from single realizations of the random field in case 1. We assume that the random field is isotropic, i.e. $d = d_h = d_v$, and has symmetric marginal distributions such that $A_i = (-\infty, -a] \cup [a, \infty), i = 1,\ldots,p$. Thus, we have five parameters to estimate: $\mu$, $\sigma^2$, $\gamma$, $d$, and $a$, and the log-likelihood is
\begin{align}
l(\mu, \sigma^2, \gamma, d, a; \mathbf x) =  \; & \log L(\mu, \sigma^2, \gamma, d, a; \mathbf x) \notag \\
= \; &
\log \phi_p(\mathbf x; \mu \mathbf 1, \sigma^2 \ \mathbf C)  \;
+ \sum_{i=1}^p \log \Phi_1(A_i; - \frac{\gamma}{\sigma}  (\mathbf x_i - \mu ), 1-\gamma^2 ) \notag \\
&- \log \Phi_p(A; \mathbf 0, (1-\gamma^2) \mathbf I_p + \gamma^2 \mathbf C)
, \label{eqn:param_likelihood}.
\end{align}
where $\mathbf C$ is a function of $d$ as previously defined, and the set $A$ is parameterized by $a$.  The restrictions on the parameters are in addition to $\sigma^2,d ,a> 0$, also $0 \leq \gamma \leq 1$, due to symmetry with respect to $\gamma$ in $l(\mu, \sigma^2, \gamma, d, a; \mathbf x)$ caused by the symmetry of $A_i$ around $0$.

The last term in Expression \ref{eqn:param_likelihood}, $\log \Phi_p(A; \mathbf 0, (1-\gamma^2) \mathbf I_p + \gamma^2 \mathbf C)$, is challenging to calculate. In order to estimate the Gaussian cdf we follow  \citet{Genz1992} and \citet{Genz2009} and use a Monte Carlo importance sampling method. By using the same set of uniform random variables for each likelihood function evaluation we ensure that the approximated likelihood is smooth; thus we are able to use standard optimization routines to optimize a Monte Carlo approximated likelihood. The algorithm is summarized in Appendix \ref{sec:app_is}, and the algorithm is implemented in C. The information matrix becomes singular as the coupling parameter $\gamma$ approaches zero \citep{Azzalini1985,Azzalini1999}, therefore we begin the optimization procedure with some steps by the derivation free Nelder-Mead simplex method, 
followed by the interior-reflective Newton method in MATLAB. 

The parameterization of selection Gaussian random field is complicated. There may exist several parameterizations that gives about the same properties in the random fields, thus the likelihood function may be multi-modal. In order to identify the global optimum we start the optimization at multiple points and choose the values of the parameters that maximize the likelihood function. In our simulation study it appeared as our approach handled singular information matrix problems and problems regarding mode identification well.  

A similar estimation procedure is used in \citet{Rimstad2012}, where also the error from the likelihood approximation is evaluated. \citet{Rimstad2012} shows that the errors in the model parameter estimates caused by the likelihood approximation are usually unproblematic when the number of Monte Carlo points $N$ is high. In this study we use $N=5\; 000$, which according to \citet{Rimstad2012} should be sufficiently high. To evaluate the likelihood takes about one minute on a regular laptop computer for a $(32 \times 32)$ grid random field.

\begin{figure}
 \centering
 \includegraphics[width=0.7\textwidth]{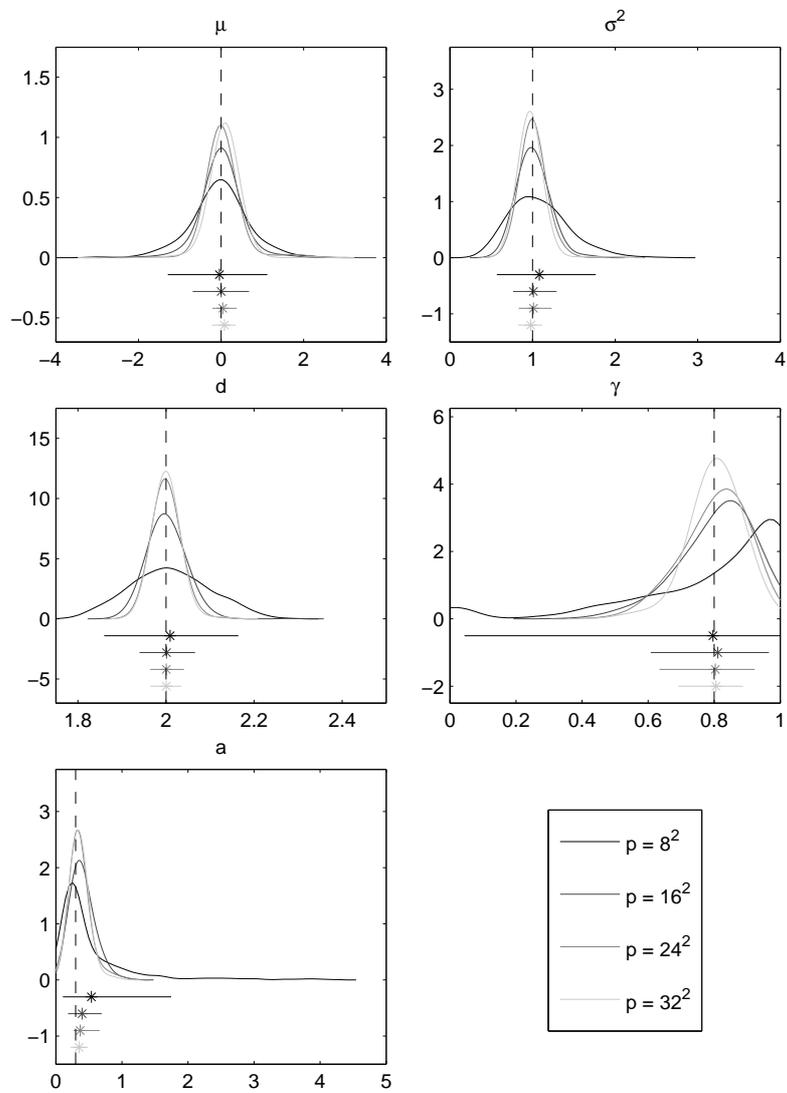}
 \caption{Density plots of parameter estimates with increasing size of the known random field. Below are means and 90\% confidence intervals, and true values as vertical dashed lines. The size range of the known random field is $p = 8^2$ to $p = 32^2$.}
 \label{fig:para_est}
\end{figure}

In order to evaluate the estimation procedure we estimate $1 \; 000$ sets of maximum likelihood parameters from $1 \; 000$ different realizations of random fields from case 1. Figure \ref{fig:para_est} displays  the distribution of $1 \; 000$ sets of maximum likelihood parameter estimates.  We let the size $p$ of the observed random field vary from $8^2$ to $32^2$. From the results in Figure \ref{fig:para_est} we see that the maximum likelihood estimates are not unbiased, but the estimators appear as consistent since the biases and variances tend toward zero with increasing size of the random field $p$. Note that the boundary values at $\gamma = 0$ and $\gamma=1$ also are acceptable values, as they represent a Gaussian and a truncated Gaussian random field, respectively.

\begin{figure}
 \centering
 \includegraphics[width=1.0\textwidth]{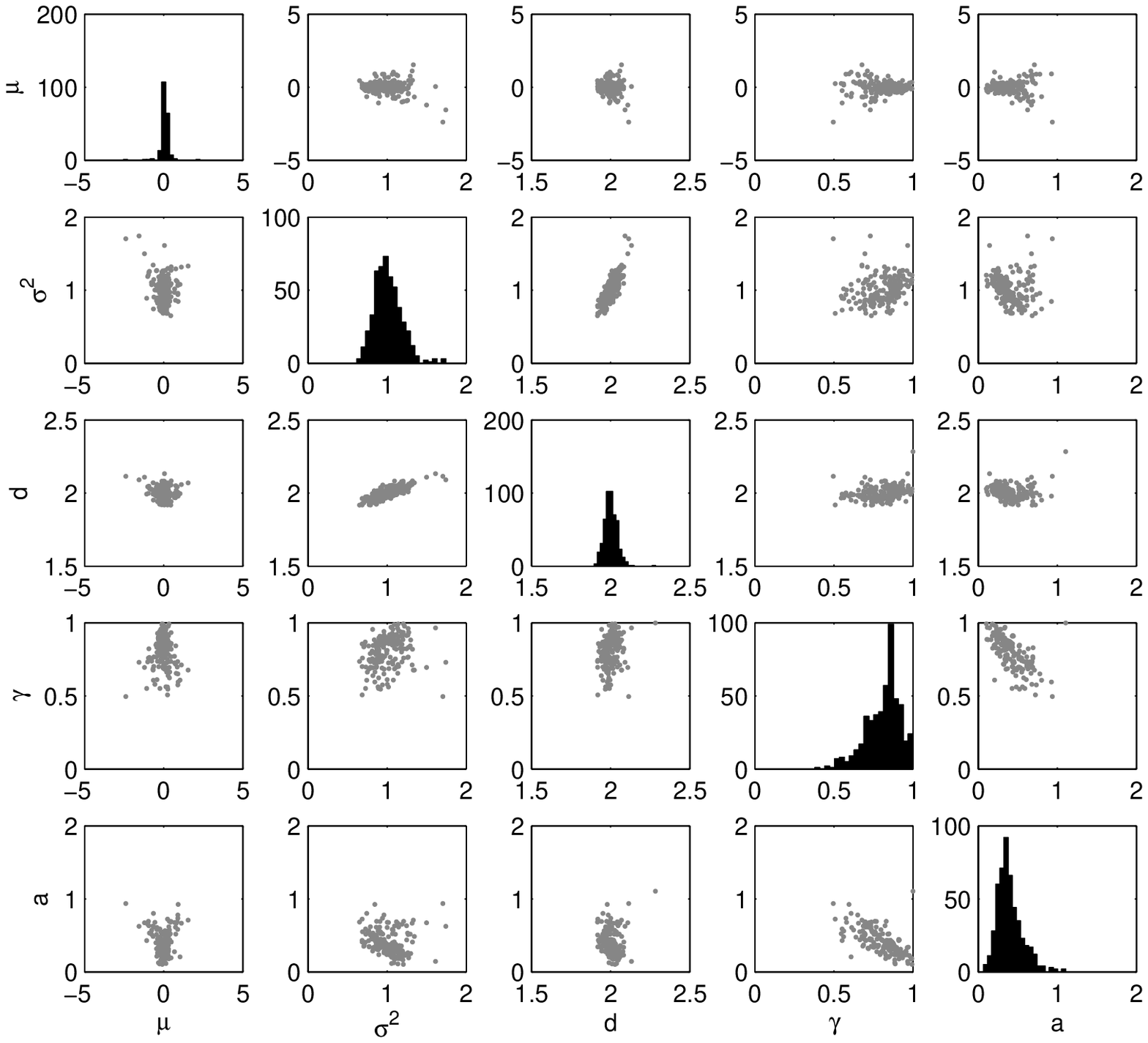}
 \caption{Cross-plot of the estimated parameters for size $p = 16^2$. }
 \label{fig:para_est_cross}
\end{figure}

Figure \ref{fig:para_est_cross} displays a cross-plot of the estimated parameters from the $1000$ realizations for $p=16^2$. The correlation between $\gamma$ and $a$ is obvious. This indicates that high values of $\gamma$ and low values of $a$ may cause similar realizations as lower values for $\gamma$ and higher values for $a$. There is also correlation between $\sigma^2$ and $d$, which may have a similar interpretation.

Case 2 through 6  have to be parameterized by models containing more model parameters. This will complicate the evaluation of the likelihood function since more ambiguities may occur. This ambiguity topic is not considered further in this study.

\section{Prediction}

\begin{figure}
 \centering
 \includegraphics[width=0.9\textwidth]{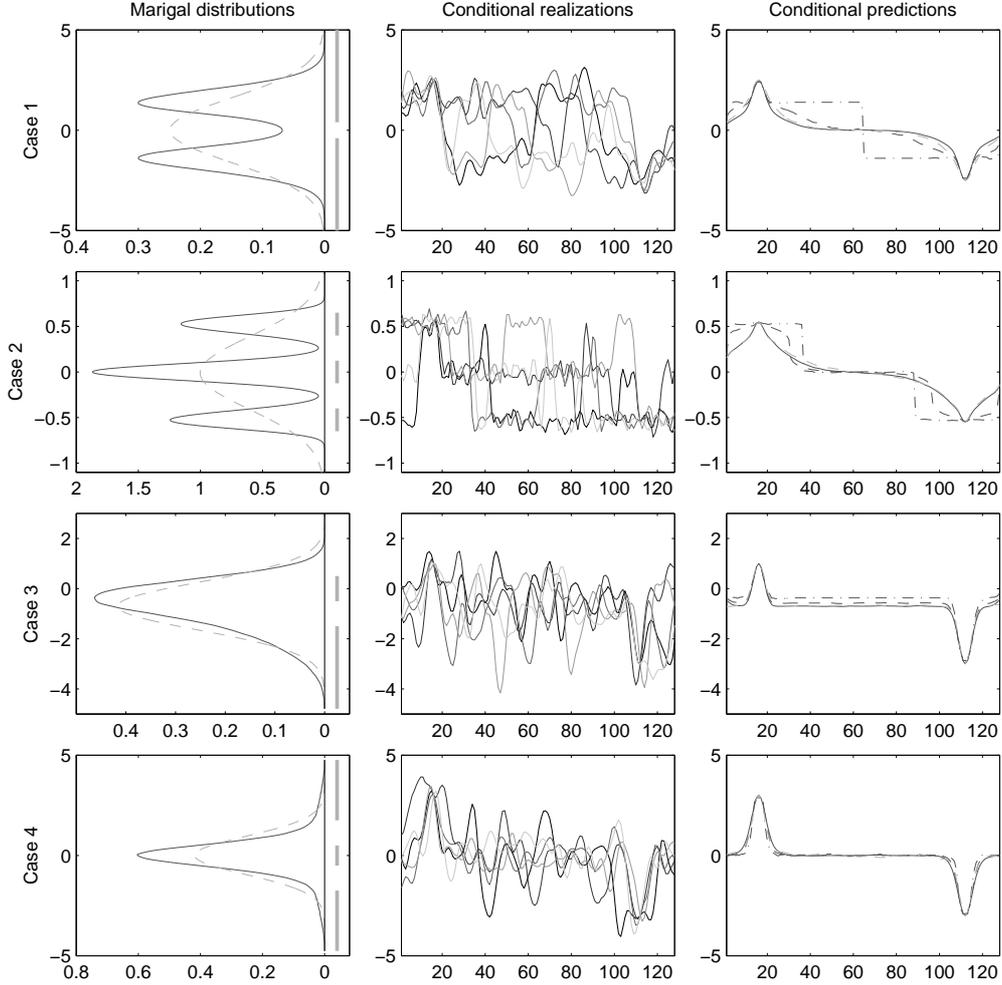}
 \caption{First column: marginal distribution before conditioning in solid black and normal distribution in dashed gray, and feasible set on latent random field on axis in solid gray. Second column: five realizations of the conditional selection Gaussian random field. Third column: conditional selection Gaussian random field predictions, with mean predictor in solid black, median predictor in dashed black, and mode predictor in dashed-dotted black. The Gaussian random field predictor (mean/median/mode) is in dashed gray.}
 \label{fig:synthetic_prediction}
\end{figure}

\begin{table}
\begin{center}
\begin{small}
\begin{tabular}{c|ccccc }
Case & $\gamma$ & $d_h$ & $A_i$ & description & cond. values\\ \hline
1    & 0.900  & 4   & $(\infty, -0.4] \cup [0.4, \infty)$ & sym. bimodal & $\pm 2.5$\\ 
2    & 0.999  & 4   & $[-0.65, -0.4] \cup [0.12, 0.12] \cup [0.4 0.65]$    & sym. trimodal & $\pm 0.55$\\
3    & 0.600  & 4   & $(\infty, -1.5] \cup [-0.5, 0.5)$ & asym. unimodal & $1.0, -3.0$\\
4    & 0.700 & 4   & $(\infty, -1.75] \cup [-0.5, 0.5] \cup [1.75, \infty)$& sym. heavy tailed & $\pm 3.0$\\
\end{tabular}
\end{small}
\end{center}
\caption{Model parameters for four predictive cases, with $\mu = 0 $ and  $\sigma^2=1$ for all cases.}
\label{tbl:synthetic_param_prediction}
\end{table}

In this section we use the selection Gaussian random field in a predictive setting. We consider a 1D random field represented on a grid of size $128$ termed $\mathbf X$, and condition on exact observed values at grid $16$ and $112$. The selection normal distribution is closed under conditioning \citep{Arellano-Valle2006a}. Thus; the predictive distribution of $\mathbf X$ given exact observed values at $16$ and $112$ is also a selection normal distribution, and it can be assessed by simulation in the same way as in the previous section by using the algorithm in Appendix \ref{sec:app_sample}.

We consider four different selection Gaussian random fields and the parameter values are summarized in Table \ref{tbl:synthetic_param_prediction}. The four cases are 1D random fields shearing about the same characteristics as case $1, 4, 5,$ and $6$ in Table \ref{tbl:synthetic_param}. We compare the selection Gaussian random field predictions to Gaussian random field predictions. The parameters in the normal distribution used for Gaussian predictions are estimated empirically from realizations of the selection Gaussian random field.

The predictions are displayed in Figure \ref{fig:synthetic_prediction}. The first column in Figure \ref{fig:synthetic_prediction} displays the unconditioned marginal distribution in location $64$ in the random field together with a marginal normal distribution, where both marginal distributions have identical two first moments. As previously we have plotted the feasible set of the latent random field with gray line segments. The second column displays realizations of the conditional distributions given the values at grid $16$ and $112$. The last column displays a conditional selection Gaussian random field mean predictor, a median predictor, a mode predictor, and a traditional Gaussian mean/median/mode predictor. In this section we will be considering predictors calculated by using the selection Gaussian random field if we do not specify anything else. 

The first row, case 1, displays a symmetric bimodal random field. The marginal distribution in Figure \ref{fig:synthetic_prediction} is clearly bimodal. We condition on the values $2.5$ and $-2.5$ at grid $16$ and $112$, respectively. The realizations of the conditioned field have a evident bimodal structure. The conditional mean predictor is almost identical to the conditional Gaussian predictor, while the conditional median and mode predictors clearly deviate from the Gaussian predictor. The mode predictor has a stepwise structure and stays in the mode that is closest to the value we condition on, and the median is somewhere between the mode and mean predictor, but closest to the mean predictor.

Case 2 is displayed in the second row and is a symmetric trimodal random field. We condition on the values $0.55$ and $-0.55$ at grid $16$ and $112$, respectively. The three modes are clearly visible in both the marginal distribution and the conditional realizations. The conditional mean predictor is in this case also almost identical to the Gaussian predictor. The mode predictor has a stepwise structure with three levels, and the median predictor is in this case more close to the mode predictor than the mean predictor.

The third row, case 3, displays an asymmetric unimodal random field. We condition on the values $1.0$ and $-3.0$ at grid $16$ and $112$, respectively. The marginal distribution is obviously skewed and the conditional realizations have a skewed structure. The mean predictor and the Gaussian predictors are again almost identical. The mode and median predictors are similar to the mean predictor except that the stationary values for the mode and median are somewhat shifted relative to the mean predictor.

Case 4 is displayed in the last row which displays a symmetric heavy tailed random field. We condition on the values $3.0$ and $-3.0$ at grid $16$ and $112$, respectively. All the conditional predictions have similar shapes, but the mode, followed by the median, decays faster toward the stationary value than the mean predictor. Again the mean and the Gaussian predictions are almost identical. The fact that the mode, median, and mean predictors are not identical entails that the conditional distributions are asymmetric.

In this section the selection Gaussian random field is used in a predictive setting. We have seen that the mean, median, and mode predictors can be very different for a selection Gaussian random field, compared to a Gaussian random field where all the three predictors are identical. The predictors are particularly different for a multi-modal random field, where the mode predictor has a stepwise structure. For random field with asymmetric marginal distributions the three predictors are not identical. We have also seen that the mean predictors for the selection Gaussian random field and Gaussian random field is almost identical when the parameters for the Gaussian random field are estimated empirically from realizations from the selection Gaussian random field.

\section{Seismic data from the North Sea}

\begin{figure}
 \centering
 \includegraphics[width=0.4\textwidth]{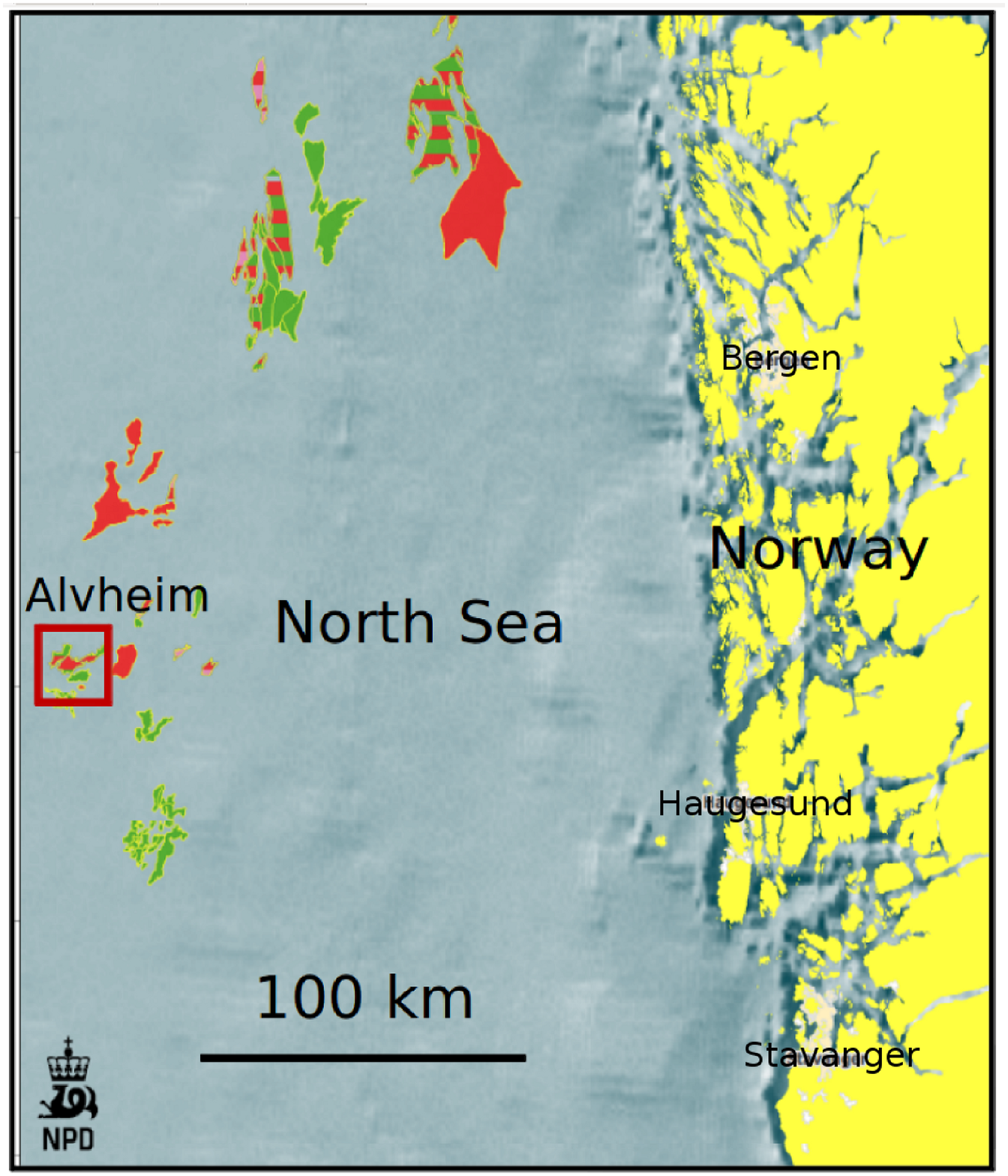}
 \caption{A North Sea map (from Norwegian Petroleum Directorate) with location of the Alvheim field with other major oil (green) and gas (red) fields.}
 \label{fig:alvheim}
\end{figure}

In this section we analyze seismic data and well observations from the Alvheim field. The Alvheim field is a turbiditic oil and gas field located on the Norwegian continental shelf in the North Sea (Figure \ref{fig:alvheim} and \Citealp{avseth:788}). The Alvheim field is buried approximately $2$ km below the sea floor. The data have previously been studied in \citet{Rimstad2011} where a Bayesian mixture model is used. In this study we use the selection Gaussian random field to model the data. We have observations from one well, and use seismic amplitude versus offset (AVO) data from one trace along this well. 

The objective of seismic AVO inversion is to invert seismic AVO data $\mathbf d$ into the logarithm of the elastic material properties $\mathbf m$.  The logarithm transformation is used to get a linear relationship between the variables of interest $\mathbf m$ and the seismic data $\mathbf d$. The elastic materials are pressure wave velocity $v_p$, shear wave velocity $v_s$, and density $\rho$. In \citet{buland:185} the problem of seismic inversion is casted in a Bayesian setting. We follow this approach; thus the posterior distribution is the objective
\begin{align}
p(\mathbf m \mid \mathbf d) = \mbox{const} \times p(\mathbf d \mid \mathbf m) \; p(\mathbf m), \label{eqn:seismic_posterior}
\end{align}
where $\mbox{const}$ is a normalizing constant, $p(\mathbf d \mid \mathbf m)$ is the likelihood, and $p(\mathbf m)$ is the prior distribution of $\mathbf m$. 

\begin{figure}
 \centering
 \includegraphics[width=0.25\textwidth]{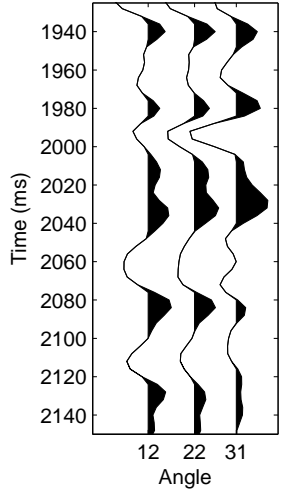}
 \caption{Seismic amplitude data in the well trace for reflection angles $12 \ensuremath{^\circ}$, $22 \ensuremath{^\circ}$, and $31 \ensuremath{^\circ}$. The depth is measured in seismic two-way traveltime.}
 \label{fig:alvheim_seismic}
\end{figure}

\begin{figure}
 \centering
 \includegraphics[width=0.7\textwidth]{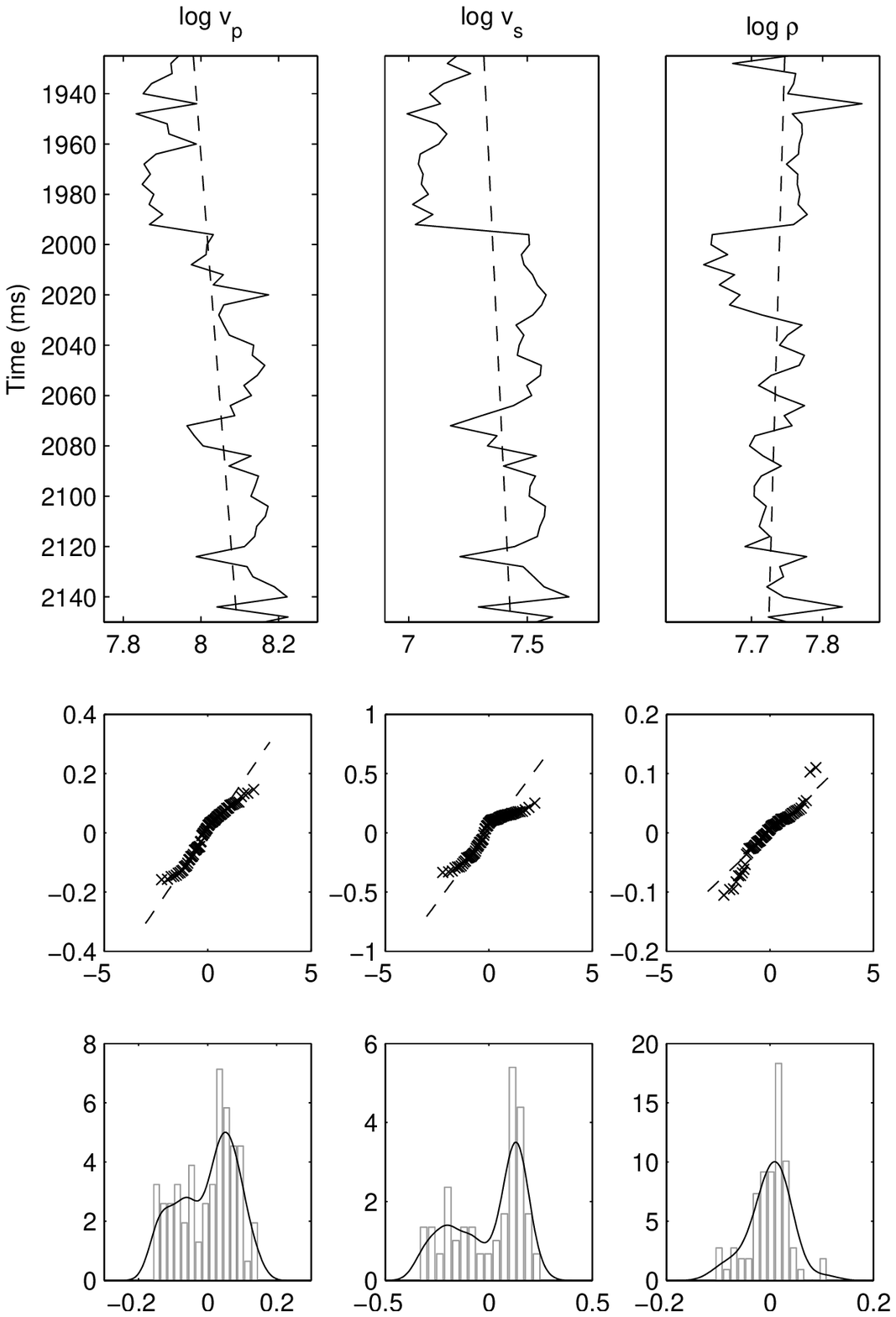}
 \caption{Well observations of logarithm of pressure-wave velocity $v_p$, share-wave velocity $v_s$, and density $\rho$. Top: elastic properties in the well with estimated linear trend in dashed black. Middle: quantile-quantile plot of residual elastic properties. Bottom: histograms and density estimates of residual elastic properties.}
 \label{fig:alvheim_well}
\end{figure}

The seismic AVO data are collected by firing air cannons on the surface and collecting the reflections from the subsurface at a set of angles. The seismic AVO data are displayed in Figure \ref{fig:alvheim_seismic}.  We have measurements for three angles  $12 \ensuremath{^\circ}$, $22 \ensuremath{^\circ}$, and $31 \ensuremath{^\circ}$ in the well trace which has length $n_t = 55$. With three angles the dimension of $\mathbf d$ is $3 \times n_t = 165$, and the dimension of $\mathbf m $ with three elastic parameters is also $3 \times n_t = 165$. The well observations $\mathbf m_w$, which are the observed values of $\mathbf m$, are displayed in Figure \ref{fig:alvheim_well}. A linear trend is estimated for each elastic parameter, and the residuals are plotted in quantile-quantile plots and histogram/density plots. The pressure-wave velocity $v_p$ and share-wave velocity $v_s$ do not fit the normal distribution assumption particularly well. The density $\rho$ has marginal distribution closer to a Gaussian and 
less deviations from the normal distribution on the quantile-quantile plot. In this study we model $\mathbf m$ by a bimodal symmetric selection Gaussian random field.

\begin{figure}
 \centering
 \includegraphics[width=0.3\textwidth]{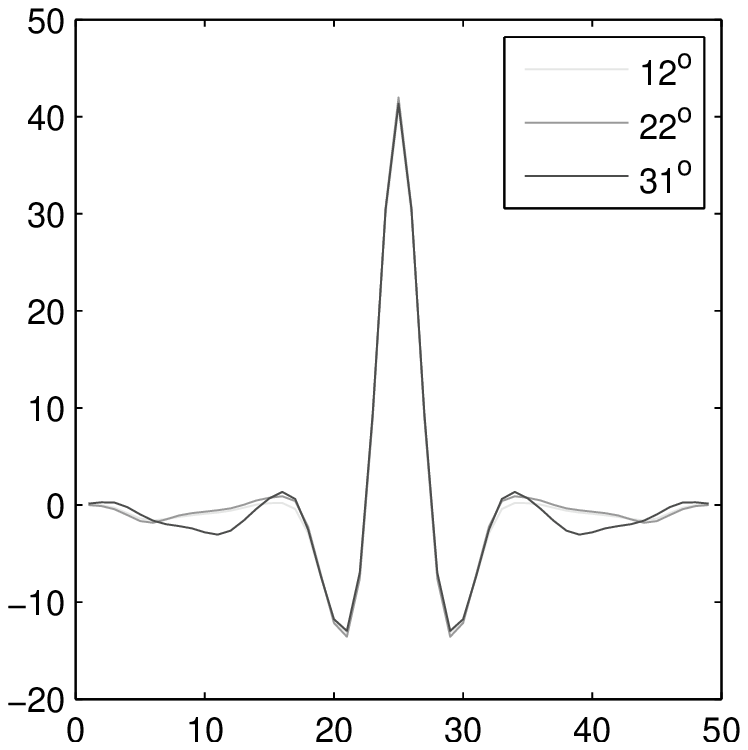}
 \caption{Seismic wavelets shape for reflection angles $12 \ensuremath{^\circ}$, $22 \ensuremath{^\circ}$, and $31 \ensuremath{^\circ}$.}
 \label{fig:alvheim_wavelets}
\end{figure}

The relation between the seismic AVO data $\mathbf d$ and the logarithm of the elastic material properties $\mathbf m$ can be modeled by a weak-contrast, convolutional, linearized Zoeppritz model \citep{akirickards,buland:185}.  The convolutional forward model is defined by $\mathbf G=\mathbf{WAD} \in \mathbf R^{3n_t \times 3 n_t}$, where $\mathbf W$ is a convolutional matrix defined by the kernels in Figure \ref{fig:alvheim_wavelets}, $\mathbf A$ is a matrix of angle-dependent weak contrast Aki-Richards coefficients \citep{akirickards}, and $\mathbf D$ is a differential matrix which calculates contrasts. The model is $\mathbf d = \mathbf G \mathbf m + \mathbf e$, where $\mathbf e$ is assumed to be a Gaussian error term with zero mean, and model 
approximation and measurement error covariance matrix $\boldsymbol \Sigma_{\mathbf e}$. The likelihood model is thus 
\begin{align}
p(\mathbf d \mid \mathbf m) = N(\mathbf G \mathbf m, \boldsymbol \Sigma_{\mathbf e}).
\end{align}
The covariance matrix is parameterized as $\boldsymbol \Sigma_{\mathbf e} = \sigma^2_{\mathbf e} \cdot \mathbf C^0_{\mathbf e} \otimes \mathbf C_{\mathbf e}$, where $\otimes$ denotes the Kronecker product, $\sigma^2_{\mathbf e}$ is the error variance, $\mathbf C^0_{\mathbf e} \in \mathbf R^{3 \times 3}$ is a wavelet correlation matrix parameterized as an  exponential correlation matrix with parameter $d^0_{\mathbf e}$, and $\mathbf C_{\mathbf e} \in \mathbf R^{n_t \times n_t}$ is a vertical correlation matrix parameterized as an  exponential correlation matrix with parameter $d_{\mathbf e}$.

The selection Gaussian random field used to model $\mathbf m$ is defined by the location parameter $\boldsymbol \mu_\mathbf m$, truncation region $A$, and the full covariance matrix for $\mathbf m$ and the truncated field:
\begin{align}
\left[
\begin{array}{cc}
\boldsymbol \Sigma^0_{\mathbf m} \otimes \mathbf C_{\mathbf m} 		
& -  \left( \boldsymbol  \Sigma^0_{\mathbf m} \left( \boldsymbol \Gamma^0 \boldsymbol \Omega^0_\mathbf m \right)^T \right) \otimes \mathbf C_{\mathbf m} \\
- \left( \boldsymbol \Gamma^0 \boldsymbol \Omega^0_\mathbf m \boldsymbol  \Sigma^0_{\mathbf m}  \right)  \otimes \mathbf C_{\mathbf m}	& (\mathbf I_3-\boldsymbol \Gamma^0)(\mathbf I_3-\boldsymbol \Gamma^0) \otimes \mathbf I_{n_t } + \left(\left(\boldsymbol  \Gamma^0 \boldsymbol \Omega^0_\mathbf m \right) \boldsymbol \Sigma^0_{\mathbf m} \left( \boldsymbol \Gamma^0 \boldsymbol \Omega^0_\mathbf m \right)^T\right) \otimes \mathbf C_{\mathbf m}
\end{array} \label{eqn:seismic_covariance}
\right],
\end{align}
where $\boldsymbol \Sigma^0_{\mathbf m} \in \mathbb{R}^{3 \times 3}$ is the covariance matrix between the three elastic material properties and $\mathbf C_{\mathbf m} \in \mathbb{R}^{n_t \times n_t}$ is a spatial exponential correlation matrix with parameter $d_\mathbf m$. The parameter $\boldsymbol \Omega^0_\mathbf m$ is a diagonal matrix with elements being the square root of the inverse elements of the diagonal matrix of $\boldsymbol \Sigma^0_{\mathbf m}$, and is used to scale the covariance matrix of the truncated field, and $\boldsymbol \Omega^0_\mathbf m \boldsymbol \Sigma^0_{\mathbf m} (\boldsymbol \Omega^0_\mathbf m)^T $ is a correlation matrix. The coupling structure is $\boldsymbol \Gamma = \boldsymbol \Gamma^0 \otimes \mathbf I_{n_t }$, where $\boldsymbol \Gamma^0 = \mbox{diag}(\boldsymbol \gamma)$, with $\boldsymbol \gamma =  [\gamma_{v_p}, \gamma_{v_s}, \gamma_{\rho}]^T$. The Expression \ref{eqn:seismic_covariance} corresponds to Expression \ref{eqn:covariance} extended to a multivariate random field, or in this case a multivariate time 
series.

The location parameter vector $\boldsymbol \mu_\mathbf m \in \mathbb{R}^{ 3 n_t }$ is parameterized with linear trends for each elastic material property and we use the trends displayed in Figure \ref{fig:alvheim_well}. The truncation region $A$ is parameterized with three parameters $\mathbf a = (a^{v_p},a^{v_s},a^{\rho})$, where we use one parameter for each elastic parameter: $A^{v_p}_i = (-\infty,-a^{v_p}] \cup [a^{v_p}, \infty)$, $i = 1,\ldots,n_t$ and similar for $v_s$ and $\rho$. The unknown parameters in the prior and likelihood models are $\sigma^2_{\mathbf e}$, $d^0_{\mathbf e}$, $d_{\mathbf e}$, $d_\mathbf m$, $\boldsymbol \Sigma^0_{\mathbf m}$, $\mathbf a$, $\boldsymbol \gamma$, which we term $\boldsymbol \theta$.

\begin{figure}
 \centering
 \includegraphics[width=\textwidth]{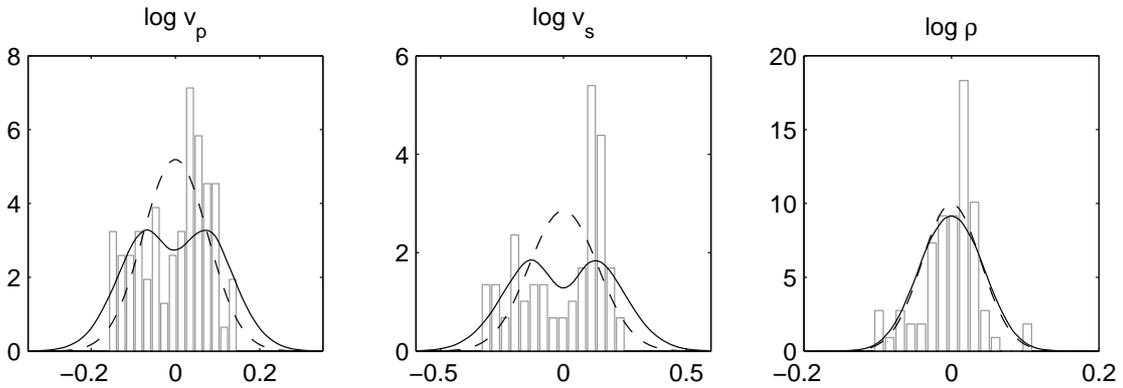}
 \caption{Estimated prior marginal models. Marginal distributions of estimated selection Gaussian random field in solid black, marginal distributions of estimated Gaussian random field in dashed black, and histograms are histograms of well observations.}
 \label{fig:alvheim_prior_select}
\end{figure}

We estimate the parameters $\boldsymbol \theta$ by using the well observations $\mathbf m_w$ and seismic observations $\mathbf d$ in the likelihood and prior model. We estimate $\sigma^2_{\mathbf e}$, $d^0_{\mathbf e}$, $d_{\mathbf e}$ by maximizing the likelihood $p(\mathbf d \mid \mathbf m_w, \sigma^2_{\mathbf e}, d^0_{\mathbf e}, d_{\mathbf e})$ with respect to $\sigma^2_{\mathbf e}$, $d^0_{\mathbf e}$, $d_{\mathbf e}$, and we estimate $ d_\mathbf m$, $\boldsymbol \Sigma^0_{\mathbf m}$, $\mathbf a$, $\boldsymbol \gamma $ by maximizing the prior $p(\mathbf m_w \mid d_\mathbf m, \boldsymbol \Sigma^0_{\mathbf m}, \mathbf a, \boldsymbol \gamma) $ with respect to $d_\mathbf m$, $\boldsymbol \Sigma^0_{\mathbf m}$, $\mathbf a$, $\boldsymbol \gamma$.  The estimated parameters are
\begin{align}
\sigma^2_{\mathbf e} = 0.402, d^0_{\mathbf e} = 7.3, & \; d_{\mathbf e} = 11.1,  \notag \\
d_\mathbf m = 1.61, 
\boldsymbol \Sigma^0_{\mathbf m}
=
\left[
\begin{array}{rrr}
    0.0073  &  0.0126 &  -0.0013 \\
    0.0126  &  0.0250 &  -0.0039 \\
   -0.0013  & -0.0039 &   0.0018
\end{array}
\right], & \;
\mathbf a = \left[
\begin{array}{r}
    0.1110 \\   0.2619 \\   0.1151
\end{array}
\right],
\boldsymbol \gamma = \left[
\begin{array}{r}
  0.8656 \\   0.9061 \\   0.3331
\end{array}
\right].
\end{align}
The entire estimation procedure takes a couple minutes on a regular laptop computer. 

We want to compare the selection Gaussian random field model to a model of a Gaussian random field. The parameter estimates for $\sigma^2_{\mathbf e}, d^0_{\mathbf e}, d_{\mathbf e}$ are the same values as for the selection Gaussian model. We obtain the Gaussian model by fixing $\boldsymbol \gamma = \mathbf 0$ in the prior model, then the unknown parameters are $d_\mathbf m$ and $\boldsymbol \Sigma^0_{\mathbf m}$. The estimated values for the Gaussian model are
\begin{align}
d_\mathbf m = 1.53, 
\boldsymbol \Sigma^0_{\mathbf m}
=
\left[
\begin{array}{rrr}
    0.0059 &   0.0093 &  -0.0007 \\
    0.0093 &   0.0195 &  -0.0025 \\
   -0.0007 &  -0.0025 &   0.0016
\end{array}
\right].
\end{align}
Figure \ref{fig:alvheim_prior_select} displays the marginal distributions of the estimated selection multivariate Gaussian random field and multivariate  Gaussian random field. The observations from the well are also displayed. The marginal distributions of the selection Gaussian random field for pressure-wave and shear-wave velocity are bimodal, and the marginal distribution of the selection Gaussian random field for density is more similar to the normal distribution, but not identical. By including more parameter we could model the asymmetry for pressure-wave and shear-wave velocity parameters, and heavy tail structure in density, but we have chosen to use a parsimonious model with only one truncation parameter in this study.

\begin{figure}
\centering
\includegraphics[width=0.6\textwidth]{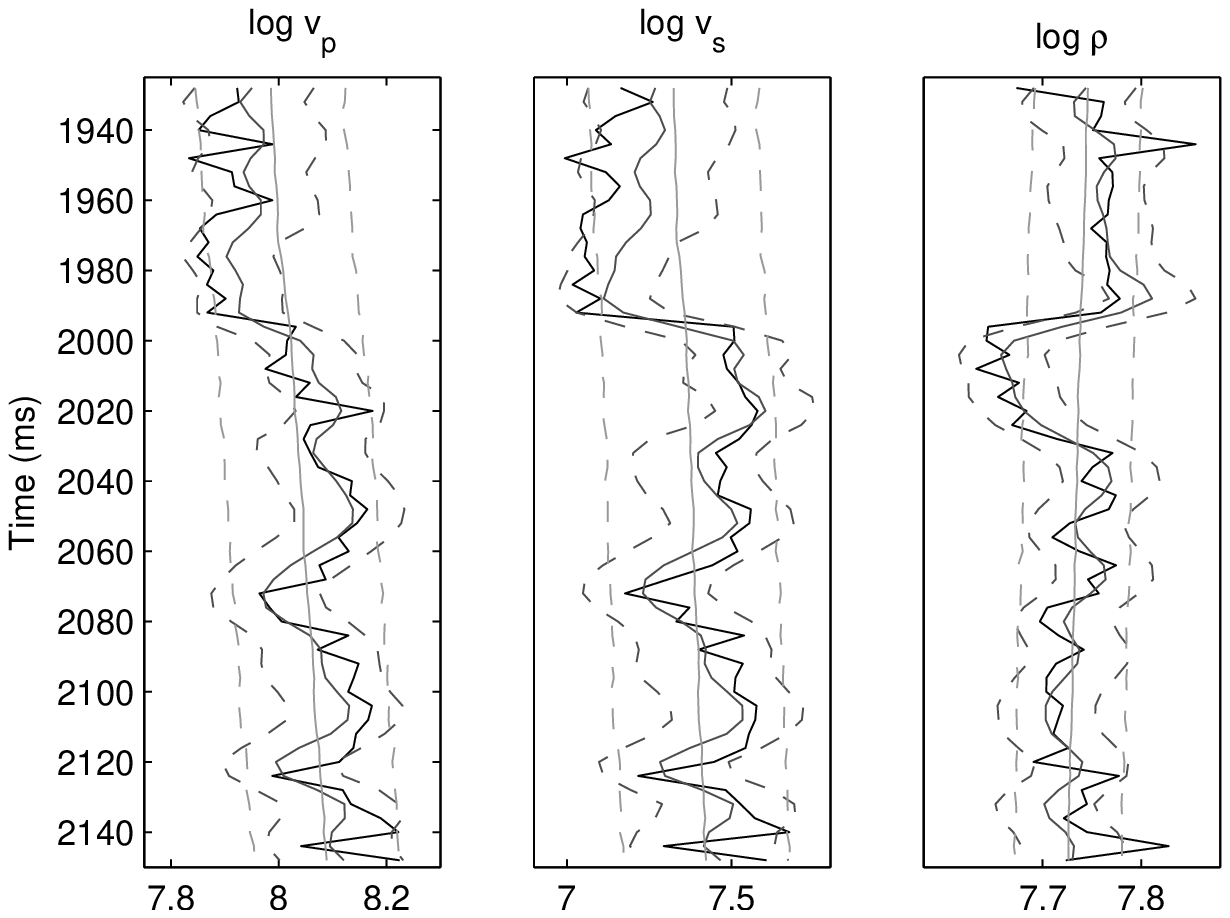}
\includegraphics[width=0.6\textwidth]{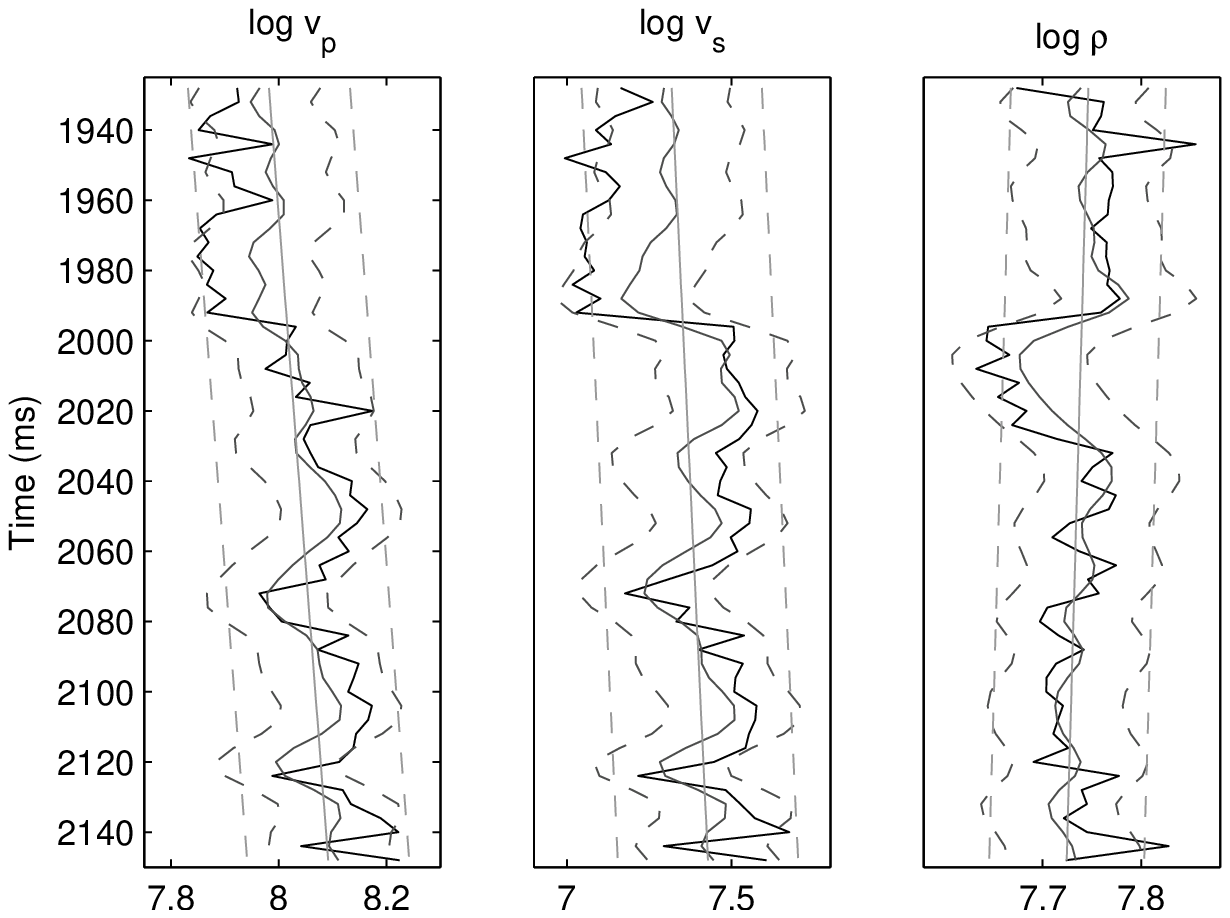}
 \caption{Well predictions.  Top: selection Gaussian model. Bottom: Gaussian model.  Well observations in solid black, posterior mean in solid dark gray, posterior 80\% prediction interval in dashed dark gray, prior mean in solid light gray, and prior 80\% prediction interval in dashed light gray.}
 \label{fig:alvheim_pred}
\end{figure}

Given the estimated parameters $ \boldsymbol{\hat \theta}$ we want to predict the elastic properties $\mathbf m$ given the seismic data $\mathbf d$ and the estimated parameters $ \boldsymbol{\hat \theta}$, and the predictive distribution is $p(\mathbf m \mid \mathbf d, \boldsymbol{\hat \theta})$, which also is a selection normal distribution due to the closure under conditioning \citep{Arellano-Valle2006a}. Note that the well log of the elastic properties $\mathbf m_w$ is only used indirectly through the estimate of $ \boldsymbol{\theta}$ in the predictive distribution. The predictive distribution is estimated by sampling $10 \; 000$ realizations using the MH algorithm in Appendix \ref{sec:app_sample}, which takes a couple of minutes on a regular laptop computer. The predictions of the elastic material in the well trace are displayed in Figure \ref{fig:alvheim_pred} for both the selection Gaussian and Gaussian model. The black solid lines are well observations, solid dark gray lines are posterior means, 
dashed dark gray lines are posterior 80\% prediction intervals, solid light gray lines are prior means, dashed light gray lines are prior 80\% prediction intervals. Predictions from the Gaussian model are not able to follow jumps in the value of the variables. The Gaussian predictions fall faster back to the prior mean value compared to the predictions by the selection Gaussian model. The median and mode predictors appear as very similar to the mean predictor for the selection Gaussian model and thus are not shown. Recall that in this example we have observations at all locations although with high observation error. The observation design is very different in the synthetic prediction cases previously presented where two exact observations are used and then the mean, median, and mode predictors appear as very different in the bimodal case.

\begin{table}
\begin{center}
\begin{tabular}{|c|rr|rr|rr|}
\hline
& \multicolumn{2}{|c|}{MSE} & \multicolumn{2}{|c|}{Prior 80\% coverage} & \multicolumn{2}{|c|}{Posterior 80\% coverage} \\
	& Selection  	& Gaussian 	& Selection	& Gaussian 	& Selection & Gaussian 	\\
\hline
$\log v_p$	& 0.0034 	& 0.0050 	& 0.84		& 0.88 		& 0.85 & 0.96		\\
$\log v_s$	& 0.0112 	& 0.0191 	& 0.82		& 0.89 		& 0.84 & 0.87		\\
$\log \rho$	& 0.0009 	& 0.0011 	& 0.82		& 0.95		& 0.83 & 0.89		\\
\hline
\end{tabular}
\end{center}
\caption{Summary of well predictions for the selection Gaussian and Gaussian model. Mean square error (MSE) of predictions, posterior and prior 80\% coverage of prediction intervals.}
\label{tbl:well}
\end{table}

The mean square errors (MSE) and prior and posterior coverages are listed in Table \ref{tbl:well}. The mean square errors for the elastic properties are reduced by about $20$-$40\%$ when we compare the selection Gaussian model with the Gaussian model. The prior 80\% coverages are a little higher than 80\% for both models, although closer to 80\% for the selection Gaussian model. The changes from the prior to the posterior coverage are smaller for the selection Gaussian model than for the Gaussian model. 

\begin{figure}
\centering
  \includegraphics[width=0.6\textwidth]{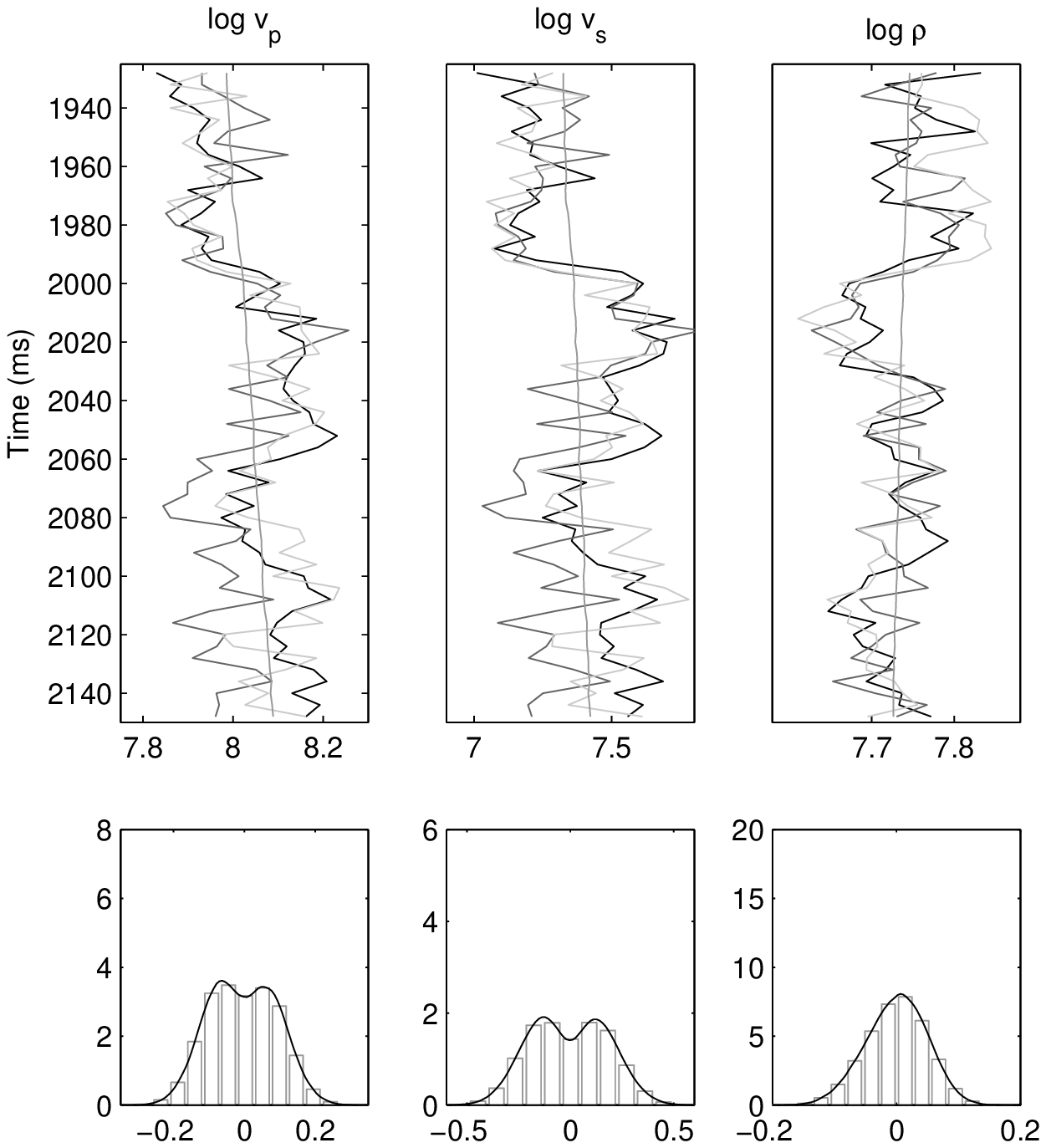}
  \includegraphics[width=0.6\textwidth]{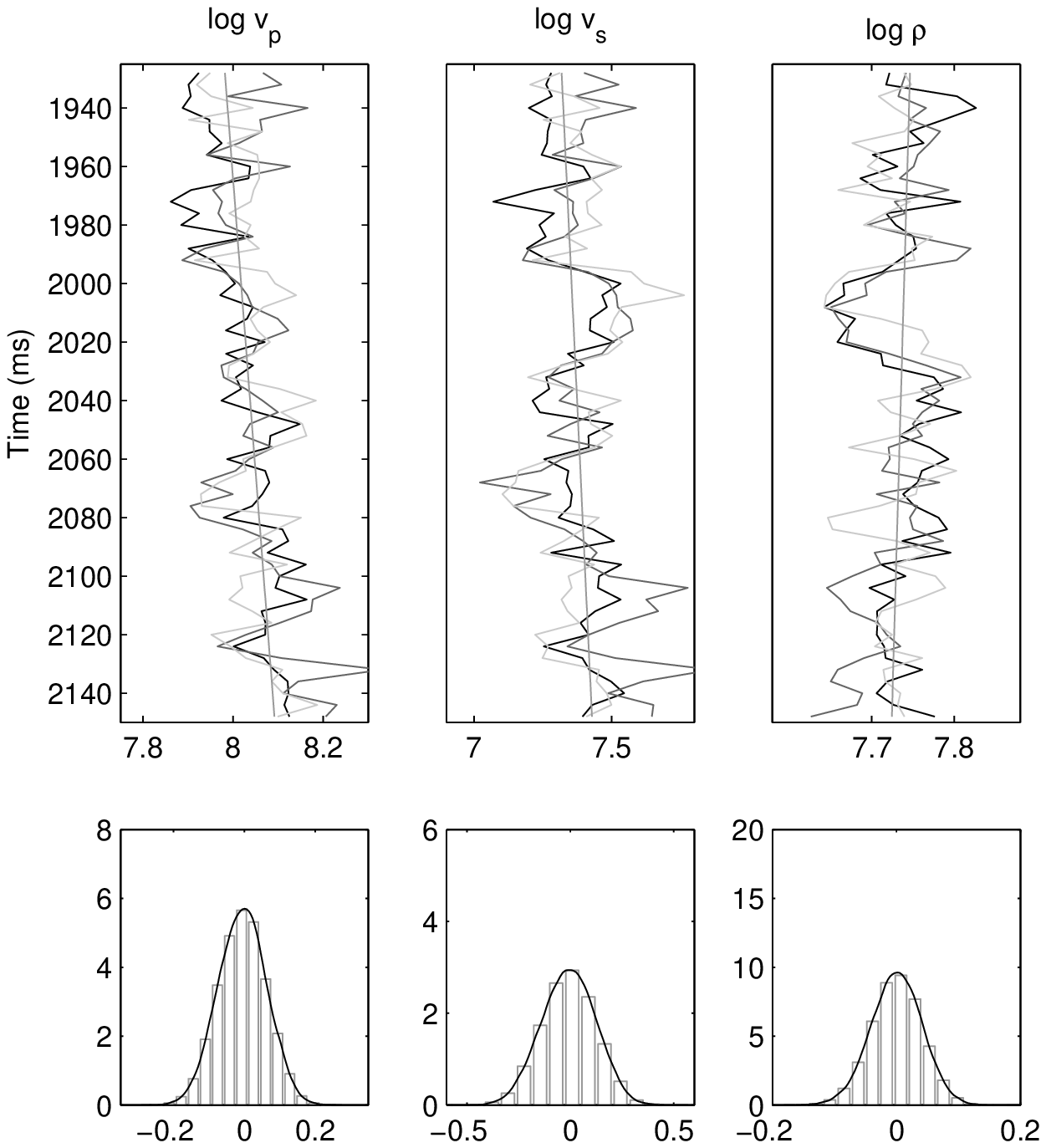}
 \caption{Three simulated realizations from posterior random fields, and realizations integrated over time. Top: Selection Gaussian model. Bottom: Gaussian model.}
 \label{fig:alvheim_pred_real}
\end{figure}

Realizations from the selection Gaussian and Gaussian posterior distributions are displayed in Figure \ref{fig:alvheim_pred_real}. The selection Gaussian model reproduce better the steep step in the value of the variables at about $2000$ ms. The marginal distribution for the selection Gaussian model is bimodal for pressure-wave and shear-wave velocity, and almost normally distributed for density, as we would expect from the prior marginal distributions in Figure \ref{fig:alvheim_well}. Note that the variance in the selection Gaussian realizations is large for shear-wave velocity in the interval $2075-2150$ ms since the prediction falls between the modes and hence realizations may move to either mode. This effect is a consequence of the bimodal structure of the selection Gaussian prior model.

\section{Concluding remarks}

In this study we define a selection Gaussian random field. The field is defined within the framework of the selection normal distribution \citep{Arellano-Valle2006a}. We have shown that skewness, multi-modality, and to some extent heavy tails in the marginal distributions can be modeled. An efficient MH-algorithm for sampling from the selection Gaussian random field is specified and a Monte Carlo approach for model parameter estimation is given. The family of selection normal distributions is closed under marginalization, conditioning, and linear transformations, which entails that conditional distributions easily can be calculated which simplifies predictions in the selection Gaussian random field.

Predictions based on either a mean, median, and mode criterion from a selection Gaussian random field model may be very different from predictions based on a Gaussian random field model. Further we have used the selection Gaussian random field as a prior model in seismic inversion of real data from the North Sea. We use a bimodal selection Gaussian random field prior model. The mean square errors in predictions are reduced by $20$-$40\%$ compared to using a standard Gaussian random field as prior model, and prediction intervals appear as more reliable.

\section*{Acknowledgments}
The research is a part of the Uncertainty in Reservoir Evaluation (URE) activity at the Norwegian University of Science and Technology (NTNU). We thank the operator of the Alvheim licenses, Marathon Petroleum Norge, and partners ConocoPhillips Norge and Lundin Norway for providing the data.

\bibliographystyle{agsm}
\bibliography{../share/ref}

\appendix

\section{Sampling from a truncated multivariate normal distribution} \label{sec:app_sample}

Consider the problem of sampling from a $n$-dimensional truncated multivariate normal distribution with unnormalized density $I(\mathbf x \in A) \times \phi_n(\mathbf x ; \boldsymbol \mu, \boldsymbol \Sigma)$, where $\mathbf x, \boldsymbol \mu \in \mathbb R^n$, $ \boldsymbol \Sigma \in \mathbb R^{n \times n}$, $A = A_1 \times \ldots \times A_n$, $A_i \subseteq \mathbb R$, $I(\cdot)$ is the indicator function, and $\phi_n(\mathbf x ; \boldsymbol \mu, \boldsymbol \Sigma)$ is the multivariate normal density distribution with expectation vector $\boldsymbol \mu$ and covariance matrix $\boldsymbol \Sigma$. In order to sample from this distribution we extend the Metropolis-Hastings algorithm in \citet{Robert:TruncGaussian} with a block independent proposal distribution:
\begin{align}
p^*(\mathbf x^a \mid \mathbf x^b)
&= \prod_{i=1}^q I( x_i^a \in A_i) \; \frac{\phi_1(x_i^a \mid \mathbf x_{1:i-1}^a, \mathbf x^b ; \boldsymbol \mu, \boldsymbol \Sigma)}{\Phi_1( x^a_i \in A_i \mid \mathbf x_{1:i-1}^a,\mathbf x^b ; \boldsymbol \mu, \boldsymbol \Sigma)}
, \label{eqn:mcmcproposal}
\end{align}
where $n_a$ is the block size, $\mathbf x^a \in \mathbb{R}^{n_a}, \mathbf x^b \in \mathbb{R}^{n-n_a}$, $\phi_1(x_i^a \mid \mathbf x_{1:i-1}^a, \mathbf x^b ; \boldsymbol \mu, \boldsymbol \Sigma)$ the conditional normal probability of $x_i^a$ given $\mathbf x_{1:i-1}^a$ and $\mathbf x^b$, and $\Phi_1( x^a_i \in A_i \mid \mathbf x_{1:i-1}^a, \mathbf x^b  ; \boldsymbol \mu, \boldsymbol \Sigma)$ is the probability of the set $A_i$ under the normal probability distribution of $x_i$ given $\mathbf x_{1:i-1}^a$ and $\mathbf x^b$. We use the notation $\mathbf x_{1:i-1} = (x_1,x_2, \ldots x_{i-1})$. The distribution in Expression \ref{eqn:mcmcproposal} is inspired by the importance sampler in \citet{Genz1992}. Note that $p^*(\mathbf x^a \mid \mathbf x^b)$ is normalized and it is easy to sample from the distribution due to the sequential structure. 

The acceptance probability in the accept/reject step is
\begin{align}
\alpha 
& = \min \left\lbrace 1, \frac{p({\mathbf x^a}' \mid \mathbf x^b)}{p(\mathbf x^a \mid \mathbf x^b)} \cdot \frac{p^*(\mathbf x^a \mid \mathbf x^b)}{p^*({\mathbf x^a}' \mid \mathbf x^b)} \right\rbrace \notag \\ 
& = \min \left\lbrace 1, \frac{ \prod_{i=1}^{n_a} \Phi_1({ x^a_i}' \in A_i \mid {\mathbf x^a_{1:i-1}}', \mathbf x^b  ; \boldsymbol \mu, \boldsymbol \Sigma)}{ \prod_{i=1}^{n_a} \Phi_1(x^a_i \in A_i \mid \mathbf x^a_{1:i-1}, \mathbf x^b  ; \boldsymbol \mu, \boldsymbol \Sigma)} \right\rbrace,
\end{align}
where ${\mathbf x^a}'$ is the new proposed state.
The Metropolis-Hastings algorithm is presented in Algorithm \ref{alg1}.

\begin{algorithm}[H]
\DontPrintSemicolon
\BlankLine\;
Initialize $\mathbf x$ with a value in $A$. \;
\textbf{Iterate} \;
\quad Choose one element $i$ at random in $\mathbf x$.  \;
\quad Find the set of the $n_a$ closest by correlation element to $i$. \; 
\quad Define the set of the $n_a$ elements $a_i$ and $b_i$ as it complement. \;
\quad Sample $\mathbf x'_{a_i \mid b_i} \sim p^*(\mathbf x^{a_i} \mid \mathbf x^{a_i})$. \;
\quad Accept $\mathbf x'_{a_i \mid b_i}$ with probability $\alpha$. \;
\textbf{End} \;
\BlankLine \;
\caption{Sampling from truncated multivariate normal distribution}
\label{alg1}
\end{algorithm}

In practice we calculate the conditional distributions in Algorithm \ref{alg1} in advance. To save memory and time we also limit the elements in $\mathbf x$, i.e. sets eligible for choice, such that all elements in $\mathbf x$ has approximately equal update probability. We normally use the block size $n_a = 100$.

\section{Monte Carlo estimation of multivariate normal probabilities} \label{sec:app_is}

Consider the problem of estimating the multivariate normal probability 
\begin{align}
\Phi_n( A;\boldsymbol \mu, \boldsymbol \Sigma)
&= \int I(\mathbf x \in A) \; \phi_n(\mathbf x;\boldsymbol \mu, \boldsymbol \Sigma) \; \mathrm d \mathbf x,
\end{align} 
where $\mathbf x, \boldsymbol \mu \in \mathbb R^n$, $ \boldsymbol \Sigma \in \mathbb R^{n \times n}$, $A = A_1 \times \ldots \times A_n$, $A_i \subset \mathbb R$, $I(\cdot)$ is the indicator function,  and $\phi_n(\mathbf x ; \boldsymbol \mu, \boldsymbol \Sigma)$ is the multivariate normal density distribution with expectation vector $\boldsymbol \mu$ and covariance matrix $\boldsymbol \Sigma$.
The usual importance sampling Monte Carlo approximation is 
\begin{align}
\Phi_n(A;\boldsymbol \mu, \boldsymbol \Sigma) 
&\approx \sum_{j=1}^N I(\mathbf x^j  \in A) \; \frac{\phi_n(\mathbf x^j;\boldsymbol \mu, \boldsymbol \Sigma)}{f_n(\mathbf x^j;\boldsymbol \mu, \boldsymbol \Sigma)},
\end{align}
with $\mathbf x^j \sim f_n(\mathbf x;\boldsymbol \mu, \boldsymbol \Sigma); \; j = 1,\ldots N$ and $N$ is the number of Monte Carlo sampling points. We extend the approach presented in \citet{Genz1992} by allowing $A_i$ to consist of several intervals, and use
\begin{align}
f_n(\mathbf x;\boldsymbol \mu, \boldsymbol \Sigma)
&= \prod_{i=1}^n I(x_i \in A_i) \; \frac{\phi_1(x_i \mid \mathbf x_{1:i-1} ; \boldsymbol \mu, \boldsymbol \Sigma)}{\Phi_1(A_i \mid \mathbf x_{1:i-1} ; \boldsymbol \mu, \boldsymbol \Sigma)},
\end{align}
as importance function, where $\phi_1(x_i\mid \mathbf x_{1:i-1}; \boldsymbol \mu, \boldsymbol \Sigma)$ the conditional normal probability of $x_i$ given $\mathbf x_{1:i-1}$, and $\Phi_1( A_i \mid \mathbf x_{1:i-1}; \boldsymbol \mu, \boldsymbol \Sigma)$ is the probability of the set $A_i$ under the normal probability distribution of $x_i$ given $\mathbf x_{1:i-1}$. We use the notation $\mathbf x_{1:i-1} = (x_1,x_2, \ldots x_{i-1})$. However, we also introduce a mean shift parameter $\boldsymbol \eta$ in the importance function which is important for asymmetric sets $A_i$. Then the importance sampling approximation appear as
\begin{align}
\Phi_q(A;\boldsymbol \mu, \boldsymbol \Sigma) 
& \approx \sum_{j=1}^N \frac{\phi_n(\mathbf x^j;\boldsymbol \mu, \boldsymbol \Sigma)}{\phi_n(\mathbf x^j;\boldsymbol \mu + \boldsymbol \eta, \boldsymbol \Sigma)} \prod_{i=1}^n \Phi_1(A_i \mid \mathbf x^j_{1:i-1} ; \boldsymbol \mu + \boldsymbol \eta, \boldsymbol \Sigma), 
\end{align}
with $\mathbf x^j \sim f_n(\mathbf x;\boldsymbol \mu+ \boldsymbol \eta, \boldsymbol \Sigma), \; j = 1,\ldots N$.

\end{document}